\documentclass[10pt]{article}
\renewcommand{\baselinestretch}{1} 
\begin{document}

\centerline
{\bf NONLINEAR BRAIN DYNAMICS AND MANY-BODY FIELD
DYNAMICS\footnote{Invited talk presented at Fr\"ohlich Centenary
International Symposium "Coherence and Electromagnetic Fields in
Biological Systems", July 1-4, 2005, Prague, Czech Republic.}}
\bigskip
\centerline{\bf Walter J. Freeman$^\dagger$ and Giuseppe
Vitiello$^{\dagger\dagger}$}

\bigskip
\centerline{$^{\dagger}$Department of Molecular and Cell Biology}

\centerline{University of California, Berkeley CA 94720-3206 USA}

\centerline{http://sulcus.berkeley.edu}
\centerline{$^{\dagger\dagger}$Dipartimento di Fisica ``E.R.
Caianiello'', INFN \& INFM}

\centerline{Universit\'a degli Studi di Salerno, Salerno, Italia}

\centerline{http://www.sa.infn.it/giuseppe.vitiello/vitiello/}

\renewcommand{\baselinestretch}{1.5} 


\vskip .5cm \centerline{\bf Abstract}
\medskip
\par \noindent
{We report measurements of the brain activity of subjects engaged
in behavioral exchanges with their environments. We observe brain
states which are characterized by coordinated oscillation of
populations of neurons that are changing rapidly with the
evolution of the meaningful relationship between the subject and
its environment, established and maintained by active perception.
Sequential spatial patterns of neural activity with high
information content found in sensory cortices of trained animals
between onsets of conditioned stimuli and conditioned responses
resemble cinematographic frames. They are not readily amenable to
description either with classical integrodifferential equations or
with the matrix algebras of neural networks. Their modeling is
provided by field theory from condensed matter physics.}

\bigskip
\medskip
\par \noindent








\bigskip
\medskip
\par \noindent



\newpage

The study of the brain functions in animal and human subjects
requires observations and measurements of the brain activity and
the formulation of dynamical models describing the observed
behavior of neural populations, axons, dendrites, glia and cell
bodies. We can describe brain functions at higher levels by using
the tools provided by classical physics and statistical mechanics,
with the associated mathematical machinery of algebraic methods
and of sets of coupled differential equations. The achievements we
have thus reached have enabled us to recognize and document the
physical states of brains; the dynamics of neurons; the functions
of membranes and organelles comprising their parts; and the
molecular and ionic ingredients that constitute the basic neural
machinery of brain function. We also observe, however, brain
states which are characterized by coordinated oscillation of
populations of neurons that are changing rapidly with the
evolution of the meaningful relationship between the subject and
its environment,  established and maintained by active perception,
which are not readily amenable to description either with
classical integrodifferential equations or the matrix algebras of
neural networks (Freeman, 2000; 2001; Vitiello, 2001). In this
paper we report indeed our  measurements of the brain activity of
subjects  engaged in behavioral exchanges with their environments
whose modeling is provided by field theory from condensed matter
physics rather than from classical dynamics.

Electroencephalographic (EEG) records were collected from 8x8
arrays of electrodes fixed on the pial surfaces of primary sensory
cortices and entorhinal cortex of rabbits and cats trained to
discriminate conditioned stimuli in the visual, auditory, somatic
and olfactory  modalities. The band pass filtered EEG signals
under the Hilbert transform gave evidence for intermittent frames
of spatial patterns of amplitude modulation of spatially coherent
carrier oscillations in the beta (12-30 Hz)  and gamma (20-80 Hz)
ranges based on long-range correlation. The events resembled
multiple overlapping frames in sequences during each act of
perception, like cinematographic representations on multiple
screens (Freeman, Burke, et al., 2003a; Freeman, 2004a). We
identify these screens with regions in the embedding medium of the
neurons, the neocortical neuropil. The large size, rapid
formation, variety of detail, and perceptual remoteness from
sensory input of normal frames requires a phase transition which
we believe cannot be explained within the context of classical
physics.

    {\it Observations of long-range correlations}

The formation and maintenance of shared oscillations by phase
transition (Freeman, 2004a,b; 2005) depends on rapid communication
among the neurons at multiple hierarchical levels. The loop
current of dendrites and axons is the chief agent for
intracellular communication, and the action potential is the chief
agent for intercellular communication from each part of the brain
to every other part. This propagating wave  is the mechanism by
which multicellular organisms greater in size than about a
millimeter overcome the limitations of diffusion as an essential
technique for distant communication. However, the length of all
but a few axons is a small fraction, only a tenth or less, (Houk,
2001) of the observed distance of long-range correlated activity,
with the requirement for synaptic renewal at each successive
relay. Even the presence of relatively sparse long axons, which
provide for high velocity jumps to 'seed' areas over long
distances cannot explain the long-distance spatial coherence with
phase dispersion at every transmission frequency, especially for
chaotic oscillations without fixed frequencies.

Further, although both electric fields and magnetic fields
accompany dendritic currents, there are no significant
electromagnetic fields (radio waves) to carry coordinating
information over large intracortical and intercortical distances,
because the electroencephalographic oscillations deriving from
fields of dendritic synaptic current are too low in frequency and
excessive in wavelength, and the electric and magnetic
permeabilities differ in the ratio of 80:1. Intensities of the
extracellular forces are 2-3 orders of magnitude smaller than the
transmembrane potential differences. Resorting to electric
potential gradients and Coulomb forces of the EEG has been shown
(Freeman and Baird, 1989) to be inadequate to account for the
long-range of the observed coherent activity, largely owing to the
shunting action of glia that reduce the fraction of extracellular
dendritic current penetrating adjacent neurons. Like the decay in
diffusion potential the fall of electric potential with distance
is too rapid, and the intensity of the Coulomb forces is too weak
to explain either the abruptness of apparent discontinuities in
the EEG phase with the onset of a phase transition or the
entrainment of the oscillation on resynchronization within the
observed time windows of 3-7 msec over distances of 1 to 19 cm
(Freeman, Ga\'al, et al., 2003; Freeman, Burke, et al., 2003a;
2003b).

   {\it Our resort to many-body field theory}

Thus, we are led to  conclude that classical tools, such as, e.g.,
classical nonlinear dynamics and classical statistical mechanics,
do not suffice. We then turn to the mathematical machinery of
many-body field theory that enables us to describe phase
transitions in distributed nonlinear media having innumerable
co-existing and overlapping ground states, actual and potential.
Indeed, many-body field theory is the only existing theoretical
tool capable to understand the dynamical  origin of long range
correlations, their rapid and efficient formation, their
stability, the multiplicity of coexisting and non interfering
ground states, their degree of coherence and ordering, the rich
sequence of phase transitions, as we observed. It is a fact that
many-body quantum field theory has been devised and constructed in
the past decades exactly to face the understanding of features
like ordered pattern formation and phase transitions in condensed
matter physics, similar to the ones we observed in the brain
dynamics, which could not be understood in the frame of classical
physics.

This does not mean that the biochemistry, the usual
neurophysiological analysis and/or any other classical tool of
investigation must be put apart. Rather, it means that the brain
studies made by using these traditional classical tools might be
further boosted by the study of the underlying microscopic
dynamics which allows to understand the richness of the observed
phenomenology.

The field theoretic model to which we resort  is the extension to
dissipative dynamics (Vitiello, 1995; 2001) of the many-body brain
model originally proposed by Ricciardi and Umezawa (1967) and
developed by Stuart, Takahashi, et al. (1978; 1979) and by Jibu
and Yasue (1992; 1995). According to such a model the production
of activity with long-range correlation in the brain takes place
through the mechanism of spontaneous breakdown of symmetry (SBS)
(Umezawa, 1993). The immediate consequence of SBS in material
substrates is the condensation of quanta, called the
Nambu-Goldstone boson particles, or waves or modes, in the least
energy state of the system (the system ground state or vacuum).
The symmetry which is considered is the electrical dipole
rotational symmetry (Del Giudice, Doglia, et al., 1985; 1986; Del
Giudice, Preparata, et al., 1988; Jibu and Yasue, 1992; 1995). As
we discuss in what follows, we believe we can follow this same
path, since the water matrix and any of the biomolecules entering
the brain cellular components are endowed with a characteristic
electric dipole moment. The input coming to the brain from the
environment is assumed to be the trigger of the breakdown of this
symmetry. Dissipation plays a crucial role since the brain is
permanently open to its environment. The dissipative dynamics is
specifically responsible of the occurrence of a multiplicity of
coexisting vacua, each one characterized by its own density of
boson condensate (Vitiello, 1995).

{\it Our bridge to nonlinear brain dynamics}

The adoption of such a field theoretic approach enables us to
model the whole cerebral hemisphere and its hierarchy of
components down to the atomic level as a fully integrated
macroscopic quantum system, namely as a macroscopic system which
is a quantum system not in the trivial sense that it is made, like
all existing matter, by quantum components such as atoms and
molecules, but in the sense that some of its macroscopic
properties cannot be described without recourse to quantum
dynamics. One of the merits of the dissipative many-body model
consists in the fact that the classicality of nonlinear, chaotic
dynamics (Kozma, Freeman, 2002; Kozma, Puljic, et al., 2004;
Skarda, Freeman, 1987; Tsuda, 2001) is derivable from it (Pessa
and Vitiello, 2003; 2004;). As observed by Atmanspacher (2004),
the dissipative quantum field model presents the further advantage
of directly addressing to the neuronal level. There are other
'quantum models', mostly formulated in the Quantum Mechanics
frame, which not only do not allow for long range correlations and
phase transitions, but also do not provide the transition to the
classical scale.

A central concern in our attempt to apply many-body physics is
then expressed in the question:  What might be the 'bridge'
between microscopic, atomic and molecular, units and the
macroscopic neural activity as we observe it? Typically, the unit
of neural activity is taken to be the action potential, the
dendritic postsynaptic potential, the chemical packet in the
synaptic vesicle, or an electric operator in a gap junction or an
ephapsis. On the one hand, the neuron, cell body, synapse,
microtubule, vesicle, electrochemical waveform, and other
microscopic structures and functions are not to be considered as
quantum objects in our analysis. The Planck constant, h, is
undeniably the unit of action at the atomic scale and below, but
it is not the decisive factor at the level of neuronal
populations. What appears to emerge from our experiments is a
"wave packet" (Freeman, 1975/2004; 2000) acting as a bridge from
quantum dynamics at the atomic level through the microscopic
behavior of classical cells to the macroscopic properties of large
populations of neurons. The wave packet we refer to is a
mesoscopic collective field of action that has measurable field
properties: the phase and the amplitude and their spatial and
temporal rates of change (gradients) at each point in the
sustaining neuropil. We stress that our wave packet is absolutely
not to be confused with the notion of wave packet describing
probability amplitudes in Quantum Mechanics (the common
denomination is only accidental). In our field-theoretic approach,
the wave packet is a collective mode, in which a myriad of action
potentials sustains a field of neural activity that we designated
formerly as an "activity density function" (Freeman, 1975/2004),
or as a "K-field" (Freeman, 2000). It gives rise to the observable
fields of amplitude and phase functions, comprising action
potentials recorded from arrays of microelectrode and dendritic
potentials recorded from arrays of mesoelectrodes as the EEG. The
wave packet or collective mode observed in our experiments turns
out to be identifiable with the so-called Nambu-Goldstone boson
wave or mode in Bose-Einstein condensates in the dissipative
many-body model (Vitiello, 1995; 2001).

In this paper, for sake of shortness, we will only very briefly
summarize the main points of our experimental observations and for
the mathematical formalism of the dissipative brain model we refer
to the original papers (Vitiello, 1995; Alfinito and Vitiello,
2000; Pessa and Vitiello, 2003; 2004) and to Vitiello (2001) for
its extensive qualitative description. The present paper can be
considered in some sense as an extended abstract of a more
detailed description of the experimental results in relation with
the dissipative many-body model which will be presented elsewhere
(Freeman and Vitiello, 2005).

{\it Five levels of neural field activity}

In our experiments we have identified the following five levels of
the activity that we describe by the dissipative many-body field
theory model:

{\it a) A basal state of symmetry.} In a human subject in deep
sleep we have observed the occasional appearance of a homogeneous
field of fluctuations with no discernible spatial, temporal or
spectral patterns. We characterize this state as one of
"symmetry". This dynamical regime is characterized by parameters
whose values do not belong to the windows of values that allow for
the symmetry breakdown. Even in the presence of an external
stimulus (provided it is below a threshold) the system "cannot"
react to it (symmetry breakdown is not possible). The system
"sleeps". External inputs at most create uninteresting
perturbations. This is the transient "vacuum" state of the
neocortex that is described by unbroken symmetry.

{\it b) Sustained deep sleep.} These transient epochs in deep slow
wave sleep are embedded in fields of spatial patterns of phase.
The fluctuations appear to result from continuous bombardment of
all areas of neocortex by other parts of the brain, including
inputs from the sensory receptors relayed mainly through the
thalamus and mainly irrelevant, because it is the work of cortex
by habituation to establish filters to mitigate the impact of such
unavoidable bombardment on cortices. The continual perturbation
gives rise to myriad local phase transitions characterized by the
conic phase gradients, which are quenched as rapidly as they are
formed, thereby maintaining the entire cortex in a robust
metastable state. The critical parameter is the mean firing rate
of neurons that is homeostatically maintained by mutual excitation
everywhere by thresholds and refractory periods. The phase cones
have no indistinguishable amplitude patterns, so we infer that
they are related to SBS with a vanishingly short-lived order
parameter.

{\it c) The awake rest state disengaged from the environment.} The
ranges of parameter allowing higher-order SBS become potentially
accessible under the influence of external weak but behaviorally
significant stimuli. The temporal phase differences appear as
patterns of coordinated analytical phase differences (CAPD), in
which each plateau of minimal phase differences is accompanied by
a peak in mean amplitude, but without discernible or reproducible
spatial amplitude modulated patterns, owing to the lack of
engagement.

{\it d) The engaged state with arousal.} The rest state evolves
into an aroused state with increased amplitude of oscillations in
the background dendritic current that accompanies incipient
engagement of the brain with the external world including the
body. There is an implicit differentiation of the set of
compatible states, which is only realized by the overt emergence
of an amplitude pattern (Freeman, Viana Di Prisco, 1986) that is
classifiable, and that arises from SBS triggered by a relevant
stimulus. We find evidence for intermittent frames of spatial
patterns of amplitude modulation of spatially coherent carrier
oscillations in the beta and gamma ranges based on long-range
correlation resembling cinematographic frames of stationary
images. The observed recurrence of patterns points to a sequence
of phase transitions. The relevant role of the dissipative
dynamics of the many-body model manifests here in the possibility
for the simultaneous overlapping of a multitude of ground states
labelled by different values of the order parameter without, or
with reduced, reciprocal interferences and preserving their
distinct autonomy. Evidence for this is found in the observed
recurrence of patterns suggesting that they overlap, largely owing
to the autonomy of the neurons in the cortical populations
(Freeman, 2005). This enables them to participate in multiple
coordinated or coherent domains simultaneously, by contributing
the covariance of only a small percentage of the total variance of
their activity to the order parameter to which they contribute and
in accordance to which they adapt their behaviours.

The description of the formation of coherent domains is a
distinctive feature of the many-body model (Del Giudice et al.,
1988; Alfinito et al., 2002). The observation of the multitude of
ground states and of coherent domains provides a strong
justification for exploring the description of nonlinear brain
dynamics in terms of the dissipative many-body model. The concept
of the boson carrier and the boson condensate enables an orderly
and inclusive description of the phase transition that includes
all levels of the macroscopic, mesoscopic, and microscopic
organization of the cerebral patterns that mediate the integration
of the animal with its environment, down to and including the
electric dipoles of all the myriad proteins, amino acid
transmitters, ions, and water molecules that comprise the quantum
system. This hierarchical system extending from atoms to whole
brain and outwardly into engagement of the subject with the
environment is the essential basis for the ontogenetic emergence
and maintenance of meaning through successful interaction and its
knowledge base within the brain. We stress once more that neuronal
cells and other macroscopic structures are by no means considered
quantum objects in our analysis. No ambiguity should be born on
this point. The SBS provides the bridge, or change of scale, from
the microscopic quantum dynamics to the macroscopic behavior of
classical cells and their constructs. The sequence of states is
reversed on the return from engagement to rest, then to sleep and
the transients with maximal disorder.

{\it e) The seizure state.} A departure from this sequence has
been observed (Freeman, Holmes, 2005;  Freeman, Holmes et al.,
2005) in the period preceding onset of a complex partial seizure
consisting of spikes at 3/s accompanied by "absence" (loss of
consciousness) and stereotypic motor automatisms. This observation
attests to the importance of long-range correlation for the
maintenance of normal metastability of the cerebral cortex. The
seizure has been simulated (Freeman, 1986) under the conditions of
a deficit of excitation and an excess of activity of inhibitory
neurons constituting regenerative, positive feedback.

{\it Description of multiple ground states with the many-body
model}

Our measurements of the spatial patterns of dendritic currents in
the primary receiving areas have repeatedly demonstrated
dependence of the patterns on the history, context, and
significance of sensory inputs, in a word on the meanings of
inputs (Freeman, 2003a; 2003b; Freeman, Rogers, 2003) and not on
their features. We are now able to conceive a hierarchy of brain
states of varying levels of symmetry and order, with the
mechanisms of spontaneous symmetry breaking by which phase
transitions occurs. We see an act of perception as having three
stages: the attentive stage of hypothesis formation in an array of
ground states that we have previously described as an attractor
landscape with multiple basins (Freeman, 2005); the testing stage
of selection by input of one of the basins among these ground
states leading to the emergence of gamma wave packets in the
neuropil of multiple primary sensory areas that express the
attractor; and the assimilation stage with emergence of adaptive
beta wave packets in the hemispheres of the forebrain by synaptic
modification with learning that modifies the attractor and its
basin.

The essence of perception is the exceedingly rapid assimilation of
the self to the changing conditions in the environment while
maintaining a historic identity across change. These operations
require that massive numbers of neurons cooperate in spatial and
temporal patterns that shift rapidly in concert with the surround.
By repeated trial-and-error each brain constructs within itself an
understanding of its surround, which constitutes its own world
that we describe as its "double" (Vitiello, 2001). It is an
"active" mirror, because the environment impacts onto the self
independently as well as reactively. The relations that the self
and its surround construct by their interactions constitute the
meanings of the flows of information that are exchanged during the
interactions. A major difficulty in seeking to verify this theory
of perception experimentally is to model the dynamics of very
large numbers of neurons without submerging their unique
identities in statistics. Only in the past two decades has the
technology become widely available for recording and measuring
large-scale images of brain activity with sufficient detail to
reveal the intrinsic structures of the activity relating to
behavior. The frontier now resides in the theory. On the one hand,
the microscopic orientation of neural networks can express the
dynamics of neurons even to the level of individual synapses, but
they do not scale up to the millions and even billions of neurons
that cooperate in perception. On the other hand, the macroscopic
approach of brain imaging using hemodynamic techniques (fMRI, PET,
SPECT), statistical mechanics, differential equations, and the
concepts of representation from artificial intelligence fail to
encompass the incredible variety of neuronal detail that is
required to match brain functions. Now it turns out that the
conflicting demands of massive numbers and intricate detail are
not unique to brain science, but have been met in a broad range of
fields in science over the past half century. It is expedient,
therefore, for brain scientists and theoretical physicists to pool
their resources and find common cause, as we have done in
describing the correspondences between brain dynamics and quantum
field theory.

\newpage

{\bf References}

\bigskip

Alfinito, E., Vitiello, G. Formation and life-time of memory
domains in the dissipative quantum model of brain, Int. J. Mod.
Phys., 2000, B14, 853-868.

Alfinito, E., Vitiello, G., Domain formation in noninstantaneous
symmetry-breaking phase transitions, Phys. Rev., 2002, B 65,
054105.

Atmanspacher H. Quantum theory and consciousness, Stanford
Encyclopedia of Philosophy, 2004;
http://plato.stanford.edu/entries/qt-consciousness/

Del Giudice, E., Doglia, S., Milani, M., Vitiello, G. A quantum
field theoretical approach to the collective behavior of
biological systems. Nucl. Phys., 1985, B251 (FS 13), 375-400.

Del Giudice, E., Doglia, S., Dilani, M., Vitiello, G.
Electromagnetic field and spontaneous symmetry breakdown in
biological matter. Nucl. Phys., 1986, B275 (FS 17), 185-199.

Del Giudice, E., Preparata, G., Vitiello, G. Water as a free
electron laser. Phys. Rev. Lett., 1986, 61, 1085- 1088.

Freeman, W.J. Mass Action in the Nervous System. New York,
Academic Press, 1975/2004

Freeman, W.J. Petit mal seizure spikes in olfactory bulb and
cortex caused by runaway inhibition after exhaustion of
excitation.  Brain Research Reviews, 1986, 11:259-284.

Freeman, W.J. Neurodynamics. An Exploration of Mesoscopic Brain
Dynamics. London UK: Springer-Verlag, 2000.

Freeman, W.J. How Brains Make Up Their Minds. New York: Columbia
University Press, 2001.

Freeman, W.J. A neurobiological theory of meaning in perception.
Part 1. Information and meaning in nonconvergent and nonlocal
brain dynamics. Int. J. Bifurc. Chaos, 2003a, 13: 2493- 2511.

Freeman, W.J. A neurobiological theory of meaning in perception.
Part 2. Spatial patterns of phase in gamma EEG from primary
sensory cortices reveal the properties of mesoscopic wave packets.
Int. J. Bifurc. Chaos, 2003b, 13: 2513-2535.

Freeman, W.J. Origin, structure, and role of background EEG
activity. Part 1. Phase. Clin. Neurophysiol. 2004a, 115:
2077-2088.

Freeman, W.J. Origin, structure, and role of background EEG
activity. Part 2. Amplitude. Clin. Neurophysiol., 2004b, 115:
2089-2107.

Freeman, W.J. Origin, structure, and role of background EEG
activity. Part 3. Neural frame classification. Clin.
Neurophysiol., 2005, 116: 1117-1129

Freeman, W.J., Baird, B. Effects of applied electric current
fields on cortical neural activity. In  Schwartz, E. ed.
Computational Neuroscience.  New York, Plenum Press. 1989, pp.
274-287.

Freeman, W.J., Burke, B.C., Holmes, M.D., Vanhatalo, S. Spatial
spectra of scalp EEG and EMG from awake humans.  Clin.
Neurophysiol., 2003a, 114: 1055-1060.

Freeman, W.J., Burke, B.C., Holmes, M.D. Aperiodic phase
re-setting in scalp EEG of beta-gamma oscillations by state
transitions at alpha-theta rates. Human Brain Mapping, 2003b,
19(4):248-272.

Freeman, W.J., Ga\'al, G., Jornten, R. A neurobiological theory of
meaning in perception.  Part 3.  Multiple cortical areas
synchronize without loss of local autonomy. Int. J. Bifurc. Chaos,
2003, 13: 2845-2856.

Freeman, W.J., Holmes, M.D. Metastability, instability, and state
transition in neocortex. Neural Networks, 2005, in press.

Freeman, W.J., Holmes, M.D., West, G.A., Vanhatalo, S. Dynamics of
human neocortex that optimize its stability and flexibility. J.
Intell. Syst., 2005, in press.

Freeman, W.J., Rogers, L.J. A neurobiological theory of meaning in
perception.  Part 5. Multicortical patterns of phase modulation in
gamma EEG.  Int. J. Bifurc. Chaos, 2003, 13: 2867-2887.

Freeman, W.J., Viana Di Prisco, G., Relation of olfactory EEG to
behavior: Time series analysis. Behavioral Neurosciences, 1986,
100: 753-763.

Freeman, W. J., Vitiello, G., Nonlinear brain dynamics as
macroscopic manifestation of the underlying many-body field
dynamics, in preparation, 2005.

Houk, J.C. Neurophysiology of frontal-subcortical loops. Ch. 4 In
Lichter, D.G., Cummings, J.L. eds. Frontal-Subcortical Circuits in
Psychiatry and Neurology. New York, Guilford Publ., 2001,  pp.
92-113.

Jibu, M., Yasue, K. A physical picture of Umezawa's quantum brain
dynamics. In R. Trappl ed., Cybernetics and System Research,
Singapore, World Scientific, 1992, p.797-804.

Jibu, M., Yasue, K.  Quantum brain dynamics and consciousness.
Amsterdam, John Benjamins, 1995

Kozma R., Freeman, W.J.  Classification of EEG patterns using
nonlinear dynamics and identifying chaotic phase transitions.
Neurocomputing, 2002, 44: 1107-1112.

Kozma, R, Puljic, M., Balister, P., Bollobás, B., Freeman, W.J.
Emergence of collective dynamics in the percolation model of
neural populations: Mixed model with local and non-local
interactions. Biol. Cybern., 2004, in press.

Pessa, E., Vitiello, G. Quantum noise, entanglement and chaos in
the quantum field theory of mind/brain states. Mind and Matter,
2003, 1, 59-79.

Pessa, E., Vitiello, G. Quantum noise induced entanglement and
chaos in the dissipative quantum model of brain. Int. J. Mod.
Phys., 2004, B18, 841-858.

Ricciardi, L. M., Umezawa, H. Brain and physics of many-body
problems, Kybernetik, 1967, 4, 44-48. Reprint in Globus, G. G.,
Pribram, K. H., Vitiello, G. eds.   Brain and Being, Amsterdam,
John Benjamins, 2005, pp. 255-266.

Skarda, C.A., Freeman, W.J.  How brains make chaos in order to
make sense of the world. Behav. Brain Sci., 1987, 10: 161-195.

Stuart, C. I. J.,  Takahashi, Y. Umezawa, H. On the stability and
non-local properties of memory, J. Theor. Biol., 1978, 71,
605-618.

Stuart, C. I. J., Takahashi, Y. Umezawa, H. Mixed system brain
dynamics: neural memory as a macroscopic ordered state, Found.
Phys., 1979,  9, 301-327.

Tsuda I. Towards an interpretation of dynamic neural activity in
terms of chaotic dynamical systems. Behav. Brain Sci., 2001,
24:793-810.

Umezawa, H. Advanced field theory: micro, macro and thermal
concepts. New York, American Institute of Physics, 1993.

Vitiello, G. Dissipation and memory capacity in the quantum brain
model, Int. J. Mod. Phys., 1995, B9, 973-989.

Vitiello, G. My Double Unveiled. Amsterdam, John Benjamins, 2001.

\end{document}